\documentclass{tcibook}
\usepackage{fancyhea}
\usepackage{work}
\usepackage{bm}       
\usepackage{graphicx}
\usepackage{hyperref}      
\usepackage{multirow}

\newcommand{\nc}{\newcommand}  

\def\beq{\begin{equation}}
\def\eeq#1{\label{#1}\end{equation}}
\def\eeqn{\end{equation}}

\newenvironment{Eqnarray}%
   {\arraycolsep 0.14em\begin{eqnarray}}{\end{eqnarray}}
\def\beqa{\begin{Eqnarray}}
\def\eeqa#1{\label{#1}\end{Eqnarray}}
\def\eeqan{\end{Eqnarray}}

\nc{\ra}{\rightarrow}  
\nc{\slsh}{\slash\hspace*{-0.22cm}}
\def\Re{{\cal R \mskip-4mu \lower.1ex \hbox{\it e}\,}}
\def\Im{{\cal I \mskip-5mu \lower.1ex \hbox{\it m}\,}}

\nc{\vev}[1]{ \left\langle {#1} \right\rangle }
\nc{\bra}[1]{ \langle {#1} | }
\nc{\ket}[1]{ | {#1} \rangle }
\nc{\fb}{\,{\rm fb}^{-1}}
\nc{\ev}{{\rm eV}}
\nc{\kev}{{\rm keV}}
\nc{\Mev}{{\rm MeV}}
\nc{\gev}{{\rm GeV}}
\nc{\tev}{{\rm TeV}}
\nc{\mev}{{\rm MeV}}

\def\del{\partial}
\def\Dslash{\not{\hbox{\kern-4pt $D$}}}
\def\dslash{\not{\hbox{\kern-2pt $\del$}}}
\def\pslash{\not{\hbox{\kern-2pt $p$}}}
\def\ETmiss{ \not{\hbox{\kern-4pt $E$}}_T }

\def\msb{{\bar{\ssstyle M \kern -1pt S}}}

\begin{document}

\def\bibname{References}
\bibliographystyle{plain}

\raggedbottom

\pagenumbering{roman}

\parindent=0pt
\parskip=8pt
\setlength{\evensidemargin}{0pt}
\setlength{\oddsidemargin}{0pt}
\setlength{\marginparsep}{0.0in}
\setlength{\marginparwidth}{0.0in}
\marginparpush=0pt


\pagenumbering{arabic}

\renewcommand{\chapname}{chap:intro_}
\renewcommand{\chapterdir}{.}
\renewcommand{\arraystretch}{1.25}
\addtolength{\arraycolsep}{-3pt}

\thispagestyle{empty}
\begin{centering}
\vfill

{\Huge\bf Planning the Future of U.S. Particle Physics}

{\Large \bf Report of the 2013 Community Summer Study}

\vfill

{\Huge \bf Chapter 7: Underground Laboratory Capabilities}

\vspace*{2.0cm}
{\Large \bf Convener: M. Gilchriese}

\pagenumbering{roman}

\vfill

{\large  Study Conveners: M. Bardeen, W. Barletta, L.~A.~T.~Bauerdick, R. Brock,
D.~Cronin-Hennessy, M.~Demarteau, M.~Dine, J.~L. Feng, M. Gilchriese,
S. Gottlieb, J.~L.~Hewett, R. Lipton, H.~Nicholson, M.~E. Peskin,
S. Ritz, I.~Shipsey, H. Weerts}\\
\vspace{1cm}

{\large Division of Particles and Fields Officers in 2013:
J.~L. Rosner (chair), 
I. Shipsey (chair-elect), 
N. Hadley (vice-chair),
P. Ramond (past chair)}\\
\vspace{1cm}

{\large Editorial Committee:
R.~H. Bernstein,
N. Graf,
P. McBride,
M.~E. Peskin,
J.~L. Rosner,
N.~Varelas,
K. Yurkewicz}

\vfill

\end{centering}

\newpage
\mbox{\null}

\vspace{3.0cm}

{\Large \bf Authors of Chapter 7:}

\vspace{2.0cm}
 {\bf  M. G. Gilchriese},
P.~Cushman, K.~Heeger, J. Klein, K.~Scholberg, H.~Sobel, M.~Witherell

 \tableofcontents
\newpage

\mbox{\null}

\newpage

\pagenumbering{arabic}


\setcounter{chapter}{6}

\chapter{Underground Laboratory Capabilities}
\label{chap:ceo}

\begin{center}\begin{boldmath}


\begin{center}

\begin{large} {\bf Convener:  M. G. Gilchriese}\end{large}

P.~Cushman, K.~Heeger, J. Klein, K.~Scholberg, H.~Sobel, M.~Witherell
\end{center}



\end{boldmath}\end{center}

\section{Introduction}

Many of the most compelling scientific issues in particle physics can only be addressed with experiments operating at underground facilities. For example, large underground facilities are needed for direct searches for dark matter and neutrinoless double-beta decay ($0\nu\beta\beta$), and the scale and complexity of these experiments will continue to increase for the foreseeable future. The importance of underground experiments using solar, reactor, atmospheric, supernova neutrinos and neutrinos from accelerators to elucidate neutrino properties is also expected to grow over the next decade and beyond.

Underground facilities are located in North America, Europe, Asia and 
Antarctica (ice). In the last few years, new underground facilities have become operational in Canada (SNOLAB), China (China JinPing Deep Underground Laboratory --  CJPL), Spain (Canfranc Laboratory, or Laboratorio Subterr\'aneo de Canfranc -- LSC) and in the United States (Sanford Underground Research Facility --  SURF). Experiments continue to be operated or assembled at older underground facilities in Asia, Europe, the United States and in ice at the South Pole. 
The world-wide particle physics community plans to expand underground capabilities over the next few years. Significant expansions are underway or planned in China (CJPL and JUNO), Korea (RENO50 and Yangyang Laboratory), Japan (Hyper-K), France (Modane extension), India (India Neutrino Observatory -- INO), South America (Agua Negra Deep Experiment Site -- ANDES) and possibly in Finland (Center for Underground Particle Physics -- CUPP). If all of these plans are realized, general-purpose space for underground experiments would roughly double by the end of the decade. There would be major new facilities for reactor experiments at moderate depths (JUNO and RENO50) and for a new class of very large facilities for long-baseline and atmospheric neutrino experiments, proton decay and other physics.  New experiments, which would require substantial facility improvements, are also proposed at the South Pole. 

Plans for expansion of underground facilities in the United States are less developed. Currently, there are no approved plans with federal funding for expansion of underground capabilities at the Kimballton Underground Research Facility (KURF), the Soudan Underground Laboratory, the 
Sanford Underground Research Facility (SURF), or the Waste Isolation Pilot Plant (WIPP). The Long-Baseline Neutrino Experiment (LBNE) has provisional approval to be located on the surface at SURF, but design work is underway in anticipation of achieving a global collaboration to allow LBNE to be sited deep underground at SURF. 

The first-generation experiments in direct dark matter detection (G1) are those operating and producing results by 2013. G2  is  the generation of experiments under construction now or in the near future, and G3 the larger experiments being considered for construction late in this decade. All of the next-generation (G2) dark matter experiments can be accommodated by existing or planned underground facilities, assuming no reduction in these facilities for the rest of the decade. Most G2 experiments are at facilities outside the U.S. \ U.S. physicists are participating in most of the G2 experiments, and they started and are leading many of them. A G3 experiment is likely to be 5--10 times the volume of the G2 experiment of similar technology and mass reach. It seems likely that a facility with depth $\geq$ 3400 mwe (e.g., LNGS) will have sufficient depth for a G3 experiment but improved simulations and results from G2 experiments will be definitive. The U.S. does not now have an underground hall large and deep enough to house a large G3 experiment. Such a new hall in the U.S. would most naturally be at SURF. It is premature to develop plans for facilities dedicated to a large directional experiment.

Several $0\nu\beta\beta$ experiments are already under construction at existing underground facilities, all but one of these outside the U.S. U.S. involvement is currently strong in many of these experiments. Next-generation (``tonne scale'') neutrinoless double-beta-decay experiments can likely be accommodated by existing and planned facilities, but may face competition for space from G2/G3-scale dark matter experiments. One next-generation experiment with large U.S. involvement (which may or may not be sited within the U.S.), with potential participation in others, is the current U.S. planning, under the auspices of DOE nuclear physics. Depth requirements for tonne-scale $0\nu\beta\beta$ experiments depend on the choice of technology and are not yet entirely known. New information may be available on a 6-month to 2-year timescale. The path beyond tonne-scale experiments is not well defined but may require expanding underground facilities.

There are worldwide efforts towards reactor experiments at medium baseline ($\sim$50 km) and short baseline ($\sim$10~m).  Detectors for reactor experiments at $>$100~m baseline require medium-depth underground laboratories (several hundred mwe overburden). There has been strong U.S. involvement in recent reactor experiments overseas (KamLAND, Daya Bay, Double Chooz). Planning and R\&D towards future reactor experiments overseas is underway with funding commitments from the host countries (RENO-50, JUNO). There may be U.S. involvement in these experiments. There may be synergistic efforts with nuclear non-proliferation activities in the U.S. In this regard, a new U.S. remote reactor monitoring initiative requires a 500-5000 mwe site to demonstrate sensitivity to reactor antineutrinos using a large Gd-water-Cherenkov detector.  The 1600 mwe Fairport mine near Cleveland Ohio and the 2800 mwe Boulby mine in England are viable options.  A kiloton-scale device will have world-class supernova sensitivity. Upgrading to liquid scintillator may enable geo-antineutrino measurements.

There is an international effort to search for CP violation in the lepton sector. A massive detector in a neutrino beam is required. The search for nucleon decay is one of the most important topics in particle physics. Atmospheric neutrinos, observable in a large underground detector, may be sensitive to all of the currently unknown neutrino oscillation parameters. Some of the same detectors could be used to advance the search for nucleon decay, the study of atmospheric neutrinos, and other physics if the detector is located underground. This is the plan for Hyper-K (Japan) and LBNO (Finland). It would be a lost opportunity if this condition cannot be satisfied with LBNE. The use of cyclotrons (or dedicated reactors) might increase the number of potential facilities for neutrino oscillation experiments in the U.S.

There is a tremendous opportunity for physics and astrophysics from detection of a supernova neutrino (SN) burst. SN detection capability typically comes “for free” if the detector is underground but is very difficult on the surface. Bursts are rare (only every $\sim$30 years): it is therefore critical to gather as much information as possible from each event. Diverse flavor sensitivity is important for physics.  The unique sensitivity of LBNE to electron flavor will be lost if the far detector is not sited underground. Broader low-energy neutrino/nuclear physics experiments (large-scale solar neutrinos, geoneutrinos, low-energy nuclear astrophysics) may require new underground spaces at existing facilities and perhaps new facilities.

Underground space should be reserved for materials assay and storage. Selection of radiopure materials for shielding and detectors is a common need. The majority of such measurements must be done underground, requiring sensitive detectors, expert personnel, and long-term storage of materials (e.g., Cu) sensitive to cosmogenic activation. Experimental needs worldwide far outstrip current assay capability. Operation as a user facility across multiple sites with existing expertise is the most efficient use of resources and personnel. In addition, underground space should be reserved for small prototype testing and generic R\&D. New experiment technologies need to go underground to validate background performance. There is enough U.S. infrastructure space for the future if existing U.S. underground labs are maintained. Substantial past agency investment and future leverage of state, university, private and other agency (e.g., non-proliferation)  funds make it cost effective and attractive to maintain these sites for smaller experiments, generic R\&D, and as elements of a centrally managed materials assay consortium.

\section{Facilities for large underground proton decay and neutrino experiments}
\label{chap:naf1}


\subsection{Physics goals}

The main physics motivations for large underground detectors are for neutrino physics --- both oscillation physics and study of astrophysical sources --- and searches for baryon number violation. These topics span event energies ranging from less than an MeV to the TeV scale.
\begin{itemize}
\item	Long-baseline neutrino oscillation: The current generation of neutrino oscillation experiments has made measurements of two mass-squared differences, $\Delta m^2_{21}$and $|\Delta m^2_{23}|$, and now all three mixing angles.  The remaining unknowns in the three-flavor picture are the mass hierarchy, the CP phase angle $\delta$, and the octant of $\theta_{23}$. Long-baseline neutrino beams of $\sim$GeV energies coupled with large detectors can address all these unknowns, and can addition improve precision on parameters and search for beyond-the-SM physics (e.g., non-standard interactions, sterile neutrinos).  Shorter-baseline pion decay-at-rest neutrinos can in addition address CP-violation physics (e.g., DAE$\delta$ALUS).
\item	Atmospheric neutrinos: The atmospheric neutrino flux covers a wide range of energies and provides neutrinos over several orders of magnitude in baseline.   Atmospheric neutrinos, observable in a large underground detector, are sensitive to all of the currently unknown oscillation parameters.
\item	Baryon number violation: To date, the search for nucleon decay has not yielded any positive evidence, but the absence of nucleon decay, now extended beyond a lifetime of 10$^{34}$ years, has provided stringent constraints that must be addressed by any proposed grand unified theory (GUT). Additional exposure will probe and constrain these models. 
\item	Solar neutrinos:  If the large underground detector has a sufficiently low energy detection threshold, then the detector could add to our knowledge of solar neutrino oscillations and potentially probe exotic physics.  Study of solar neutrinos may also improve knowledge of solar astrophysics.
\item	Supernova neutrinos: Detection of a burst of core-collapse supernova neutrinos would bring a wealth of information on both neutrino physics and astrophysics.   There may in addition be oscillation parameter information to be had (e.g., mass hierarchy) by, e.g., comparing neutrinos from the burst that have traveled through Earth matter with those that have not. 
\item	Relic supernova neutrinos: A massive underground detector would also greatly enhance the search for relic supernova neutrinos, i.e., those from supernovae throughout the history of the universe. These neutrinos are sensitive to cosmology (e.g., star formation rate) and the supernova explosion mechanism (e.g., typical neutrino temperatures) as well as exotic scenarios like neutrino decay. 
\item	Geoneutrinos:  Detectors with very low energy threshold and low levels of background can probe neutrinos from radioactive decays in the Earth, connecting to geophysics.
\item	Short-baseline oscillations: Searches for sterile neutrinos using radioactive sources coupled with large low-energy-threshold detectors have been proposed.
\item	Neutrinoless double-beta decay: Candidate isotopes may be dissolved in detector material to enable searches for neutrinoless double-beta decay in large detectors. Detection or limits on this rare process will be critical for understanding the Majorana or Dirac nature of the neutrino.	

\subsection{Large underground detectors}
In general, the same large underground detectors built for neutrino physics can be used for baryon number violation searches --- indeed, historically, the original motivation for large underground detectors was to find proton decay predicted by SU(5).    Detectors come in several types:  iron trackers, water Cherenkov detectors, liquid argon time projection chambers, and liquid scintillator detectors.  Each has pros and cons.  
\item	\textit{Water Cherenkov detectors} (e.g., Super-K, Hyper-K) have the advantage of cheap detector material making large volumes possible, and a technology proven at multi-ton scale.  Directional information is available from the Cherenkov cone, allowing multiparticle reconstruction. Events with energies from a few to hundreds of GeV are accessible.  Disadvantages are the relatively low Cherenkov light yield and limits  on reconstruction of low-energy particles due to the Cherenkov threshold.  Water Cherenkov detectors also exist in the form of long strings of photomultipliers deployed in water or ice (e.g., IceCube, Antares).  Such detectors have Mton-scale masses and sensitivity to very high-energy neutrinos but are not capable of precision reconstruction (although future upgrades, e.g., PINGU, ORCA, MICA, may mitigate this).
\item	\textit{Liquid argon time projection chamber detectors} (e.g., Icarus, LBNE) have the advantage of excellent tracking and reconstruction, enabling high efficiency for high-energy interactions and some proton decay modes (e.g, $\nu K^+$).  However LArTPCs are unproven at multi-kton scale and entail technical challenges associated with large volumes of cryogenic liquid. Low-energy physics may be challenging.
\item	\textit{Scintillator detectors} (e.g., Borexino, KamLAND, LENA) have the advantage of very high light yield, resulting in potentially very low energy thresholds and excellent energy resolution.  The technology is also well proven. The disadvantages and challenges are relatively high expense of detector material, stringent cleanliness requirements associated with low energy thresholds, and difficulty in reconstruction of high-energy multiparticle events due to isotropy of light emission.  These detectors are most suitable for low-energy physics.
\item \textit{Tracking detectors} (e.g., Soudan II, MINOS, INO, OPERA, MIND)  have the advantages of proven technology, good reconstruction capability at high energy, and ability to employ a magnetic field to sign-select interaction products.  Disadvantages are relatively high-cost detector material and low sensitivity at low energy. 
	
\end{itemize}

\subsection{Underground requirements and issues for large detectors}
While large detectors for long-baseline neutrino experiments can be located on or near the surface due to tight event timing cuts, nucleon decay detectors must be located underground to shield them from constant cosmic-ray background. Excavations for these large detectors are costly and must be situated in regions of high rock quality to minimize construction costs. Depending on the detector technology, the infrastructure requirements can include cryogenic systems, water purification systems, air conditioning, radon control, low-background construction materials, etc. Once this laboratory infrastructure is established, other experiments, such as those for direct dark matter detection and double-beta decay, which also require this underground environment, can be located there.
There are many potential sites that can house large detectors. Since overburden is not a crucial factor for long-baseline experiments, almost any surface or near surface site in the neutrino beam would be usable.  Examples here include the possible surface installation of the LBNE LAr detector and the proposed use of a water-filled mine pit in Minnesota (CHIPS).
The overburden requirement for large detectors for nucleon decay and other physics depends to some degree on the chosen technology and the specific physics topic: see Table~\ref{tab:depth}. The low-energy thresholds needed for such physics as solar neutrino and relic supernova neutrino detection require depths where the cosmic-ray muon-induced dead times and backgrounds are sufficiently reduced.

\begin{table}[h]
\centering
\begin{tabular}{|c|c|c|}
Physics & Water  & Argon\\ \hline
Long-baseline accelerator & 1000 mwe & 0--1000 mwe \\
$p\rightarrow K^+ \nu$ & $>3000$ mwe & $>3000$ mwe \\
Day/night $^8$B solar $\nu$ & $\sim$4300 mwe & $\sim$4300 mwe \\
Supernova burst & 3500 mwe & 3500 mwe \\
Relic supernova & 4300 mwe & $>$2500 mwe \\
Atmospheric $\nu$ & 2400 mwe & 2400 mwe \\ 
\end{tabular}
\caption{Estimated depth required to study various neutrino physics and proton decay processes,  with a water Cherenkov detection or with a liquid argon detector, from~\cite{Bernstein:2009ms}.}\label{tab:depth}
\end{table}

Underground sites that could potentially accommodate a new large detector include the Homestake mine in South Dakota, which will be in the LBNE beam, the Pyhasalmi mine in Finland, which could be in the proposed LBNO neutrino beam, and the Kamioka mine in Japan in the T2K neutrino beam. Other sites include the WIPP facility in New Mexico, the INO site in India, the LNGS laboratory in Italy, and the Soudan site in Minnesota. The first two have no planned neutrino beam; the latter two have beam now but will have no beam later.  Table~\ref{tab:large_det_sites} gives a summary.

We note that detectors proposed for further reactor experiments (e.g., JUNO) with significant overburden could also be considered in the large detector category.  
We additionally note that the DAE$\delta$ALUS and IsoDAR concepts could be coupled with a large inverse-beta-decay sensitive detectors.


\begin{table}[h]
\centering
\begin{tabular}{|c|c|c|c|c|c|c|}
\hline
Facility & Location  & Overburden & Past/existing & Proposed/planned & Beam & Ref.\\ 
& & (mwe) & large & large & & \\
& & & underground & underground & & \\ 
& & & detectors & detectors & & \\ \hline\hline 
Homestake & USA & 4290 & Homestake  & LBNE (LArTPC) & Future from & \cite{Lesko:2012fp}\\
                   &        &          & chlorine & & FNAL & \\ 
                   &        &          & detector & &  & \\ \hline 
Soudan & USA & 2090 & Soudan II, MINOS & & Current from & \cite{soudan}\\
            &        &           &                             & & FNAL & \\ \hline
LNGS  & Italy & 3800 & LVD, Borexino, & & Recent from & \cite{Votano:2012fr}\\
           &        &          & ICARUS, OPERA & & CERN & \\ \hline
Pyh\"asalmi & Finland & 3900 & & MEMPHYS, LENA, & Future from & \cite{laguna}\\
                    &             &           & & GLACIER, MIND   & CERN, & \\ 
                    &             &           & &                            & Protvino & \\ \hline
SNOLAB & Canada & 6010 & SNO, SNO+ & & &\cite{Smith:2012fq} \\ \hline
Kamioka & Japan & 2700 & Kamiokande, & Hyper-K  & Current and & \cite{Suzuki:2012ft}\\ 
               &           &         & Super-K, & (Tochibora & future beam &\\
              &            &         &    KamLAND             &       or Mozumi site)                     & from J-PARC & \\ 
              &            &         &                 &                            & (T2K, T2HK) & \\ \hline
INO        &  India    & 3500 &  & ICAL & &\cite{ino}\\ \hline
Baksan        &  Russia   & 850 &  BUST & BUST upgrade & & \cite{Kuzminov:2012fv}\\ \hline
Antares        &  Mediterranean    & 2500 & Antares-12 String & ORCA & &\cite{antares}\\ \hline
IceCube & South Pole & 1500 & IceCube-86 strings & PINGU, MICA & &\cite{icecube} \\ \hline\hline
\end{tabular}
\caption{Underground sites for large proton decay and neutrino detectors.}\label{tab:large_det_sites}
\end{table}

\begin{figure}[p]
\begin{center}
\includegraphics[width=0.99\textwidth]{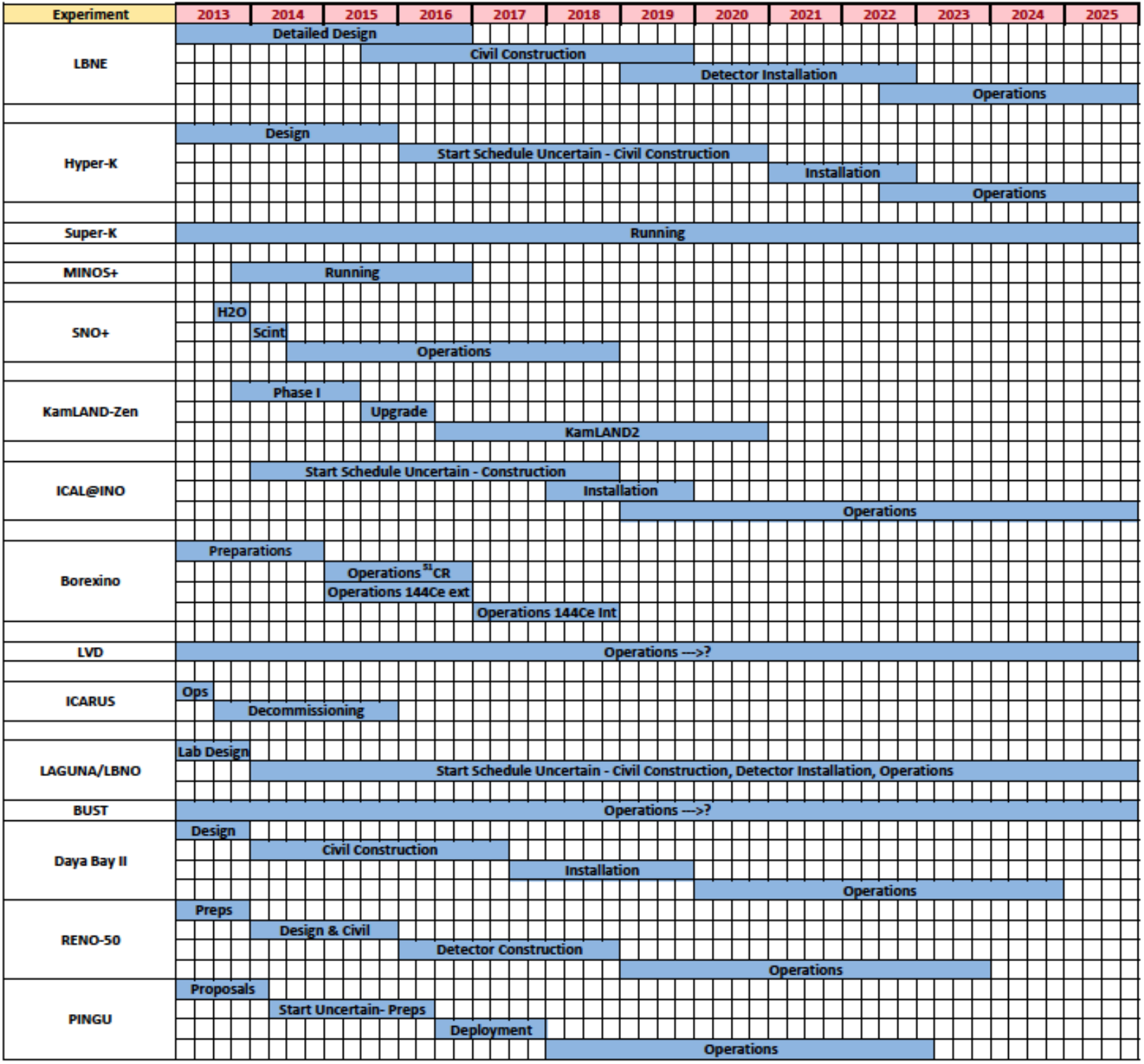}
\end{center}
\caption{Approximate timelines of underground neutrino and proton-decay experiments.}
\end{figure}

\section{Facilities and infrastructure for low-background experiments}


\subsection{Summary of requirements}

Unlike neutrino experiments, rare event searches such as dark matter and $0\nu\beta\beta$ decay place less emphasis on cavern size and more on background reduction.  Thus, they tend to prefer deeper sites with a smaller footprint, but require more in the way of radiopurity infrastructure.  Since dark matter and $0\nu\beta\beta$ decay are exploratory programs which cover large unknown parameter space, they require multiple targets and complementary techniques.  A healthy discovery program for dark matter must include a few large G2 experiments that probe favored WIMP models to the highest sensitivity; smaller experiments that explore alternate mass regimes, targets, and interaction processes; and required auxiliary efforts, such as low-background screening and storage. One needs to account for several similarly large neutrinoless double beta decay experiments worldwide that compete for the same space, but might share infrastructure such as radiopurity assay. 

Currently there appears to be enough underground space worldwide for several relatively large G2 dark matter experiments, as well as smaller complementary efforts.  The pinch points are the limited assay infrastructure worldwide and space for the future G3 dark matter experiments which can probe the irreducible neutrino background.  These G3 experiments will be physically large due to the substantial shield/veto systems required to mitigate radiogenic and cosmogenic backgrounds.  The DUSEL S4 engineering studies provide a rough envelope for the size of these experiments ($\sim$25m$\times$15m$\times$15m).     Moderate depths ($>$ 2000 mwe) can be tolerated by the use of a large, carefully designed active shield/veto, but the risk of unexpected cosmogenically-induced backgrounds is greater at shallower depths.  Sites at deeper locations also require a smaller footprint for G3, since their active shielding could be more compact.  The Cryopit and Cube Hall caverns at SNOLAB are large enough, and their 6010 mwe depth is suitable for both dark matter and $0\nu\beta\beta$ decay, but their availability on the timescale of such experiments is not clear.  The Laboratori Nazionali del Gran Sasso (LNGS) has extensive underground space capable of hosting G3 experiments at a 3000 mwe effective depth  The Laboratoire Souterrain de Modane (LSM) has a depth of 4800 mwe. A major expansion to 17,500 m$^3$ is expected to become available in roughly 2017 which could house one G3 experiment.   Jin Ping II in China will be a site of substantial depth (7000 mwe) and is expected to have 100,000 m$^3$ of space in a format large enough for G3 experiments on the 2016 timescale. The India Neutrino Observatory (INO) may have an available volume for non-neutrino experiments of about 26000 m$^3$, but there is some uncertainty about the 2018 timescale.  It should be clearly noted that there are no U.S. sites capable of hosting experiments of this size.  The available caverns at the Soudan Lab (2100 mwe) and the Sanford Underground Research Facility at Homestake (4200 mwe) are too small and would require new excavation.   

The next-generation dark matter and $0\nu\beta\beta$ decay experiments require unprecedented levels of radiopurity in their detector, target, and shielding materials.  These needs will require significant investment in the tools needed to measure such radiopurity, both new techniques to improve sensitivity and increased throughput in moderate-sensitivity production screening. While mass spectrometry and atom trap techniques can be done above ground, the vast majority of assay techniques require shielded and low-background environments to obtain the required sensitivity.  Therefore it is absolutely essential to reserve space underground dedicated to screening and associated functions like sample preparation and radon-free storage.  Space for long-term storage of activated materials, such as copper and germanium, is also beneficial.  

Space underground also needs to be reserved for R\&D, which includes small prototypes of proposed experiments, novel screening techniques, and general physics related to understanding backgrounds for experiments.   The latter projects may be measurements of muon, gamma, or neutron backgrounds as a function of geology and depth, and often include direct comparison to simulation with the purpose of improving the Monte Carlo physics models.  The reserved space should be reconfigurable, but some on-site infrastructure will enable small-scale projects that  cannot afford this overhead to realize their goals, as well as reducing the cost for larger projects.  Useful infrastructure includes gamma-ray and neutron shielding to reduce radiogenic backgrounds from the rock, an umbrella muon veto for shallower sites, cryogenic liquids or nitrogen and helium liquifiers, radon reduction systems, or even water purification and a multi-purpose water shielding tank.

\subsection{Assay needs}

Determining the number and sensitivity of the assay capabilities which should be available to the field must take account of the long lead time required for low-level assay and the screening needs preceding the commissioning of each experiment, typically by 3--5 years, as the assays inform design, as well as establish quality control of all pre-installed components. Screening needs were collected  for all dark matter collaborations by the Cosmic Frontier WIMP Direct Detection group (CF1) as part of the Snowmass process.  The results of the survey~\cite{refone} 
 indicate that the current suite of assay techniques and screeners are already inadequate for current experiments in both sensitivity and throughput. The type of assay required depends on the material to be screened, whether it is a passive or active element, and the required level of sensitivity. This implies that a variety of techniques and facilities will be necessary, which must be matched to the needs of the experiment and to the capabilities of existing screeners. 

As a function of depth and increasing sensitivity, the situation in the U.S. today is a decentralized set of detectors dedicated to specific research projects at surface sites, a few shallow depth sites which have evolved into user facilities, and a few deep sites which are challenged to provide the ultra-low counting needs for the next generation of double-beta decay, neutrino, and dark matter experiments.  These ultra-sensitive screening detectors can also be used by geology, microbiology, environmental science, and national security applications to identify radioisotopes, date samples, and measure tracers introduced into hydrological or biological systems. After the initial dialog among the different disciplines identified by the DUSEL effort, there has not been much subsequent dialogue with other fields that might also benefit from the ultra-low counting sensitivities available at deep sites. Indeed the geophysics and geomicrobiology fields have been looking for physics leadership to establish the sensitive assay capabilities.  While this is recognized as a means to enhance the user base and increase capability, it has only been possible to achieve in large national laboratories with historical ties to other communities, notably homeland security applications.

Although the background is reduced significantly by moving screening detectors underground, it is not possible to exploit extreme depths without additional effort, since backgrounds internal to the detectors themselves may become the limiting factor beyond an overburden of typically 1500 mwe.  The low-background counting facility in Gran Sasso has demonstrated its ability to achieve exceptionally low backgrounds in the detectors themselves, and consequently benefit from their additional overburden.  Moreover, for many practical applications, the counting time itself would become challengingly long.   Thus, many underground screening detectors do not benefit from placement at great depths without the commensurate attention to internal backgrounds. The use of muon veto systems can increase the effectiveness of very shallow sites such as PNNL and Oroville.

\subsection{Existing infrastructure}

Existing U.S. facilities which could be extended to users range from radiochemical and mass spectroscopy techniques on the surface to increasing levels of sensitivity moving from shallow to deep underground sites.  Table~\ref{tab:one}  provides information on assay infrastructure in the U.S.

\begin{table}[t]
\begin{center}
\begin{tabular}{|l|c|l|}
Facility &	Depth (mwe)&		Suite of detectors and technology \\ \hline\hline
Berkeley LBCF&		Surface&		2 HPGe (1 with muon veto) managed by LBNL, \\ 
   & &   \ \ \ \   100\% use by others\\ 
& & NaI, BF$_3$ counting,  shielded R\&D space \\ \hline
PNNL	&
Surface &	ICPMS: Dedicated instrument and clean room \\ & & \ \ \ \ facilities for low-background assay\\
& & 6 commercially shielded HPGe detectors  \\ 
  & &  \ \ \ \   considering use by others\\ \hline
PNNL  ULab & 	
30 &	Copper electroforming and clean machining  \\ 
& & 14-crystal HPGe array \\   & &  \ \ \ \ considering use by others \\ 
 & & multiple commercial HPGe for various stakeholders \\  \hline 
Oroville (LBNL) & 	530 & 	1 HPGe managed by LBNL \\ & & \ \ \ \  100\% use by others \\ 
& & Large shielded R\&D space \\   \hline
Kimballton (KURF)  &	1450 & 	2 HPGe managed by UNC/TUNL \\ & & 
\ \ \ \  50\% use by others \\ \hline
Soudan	&
2100 & 	1 HPGe managed by CDMS \\  & & \ \ \ \ 10\% use by others \\ 
& & 1 HPGe managed by Brown, dedicated to LUX/LZ \\
 & & 6000 m$^3$ lab lined with muon tracker  + \\ 
       & &       \ \ \ \         2 muon-correlated neutron detectors \\
 & & Large R\&D space with muon tag provided  \\ \hline
Homestake (SURF) &	4300&	1 HPGe managed by CUBED, priority to LZ, \\ 
 & & \ \ \ \  Majorana,  other users by negotiation \\ \hline
\end{tabular}

\bigskip

\caption{U.S. assay resources~\cite{refone,reftwo}.}
\label{tab:one}
\end{center}
\end{table} 

In other countries, the assay infrastructure is much more developed. It has been built up in conjunction with centralized laboratories such as SNOLAB in Canada and LNGS in Italy, and thus has the advantage of co-location with many of the experiments it services.  The existence of centers for assay creates the critical mass of experts and organization to extend the services to other fields, producing, in turn, a broad user community capable of producing a self-sustaining business model.  Tables~\ref{tab:two}  and \ref{tab:three} 
provide information on foreign assay infrastructure.

\begin{table}[p]
\begin{center}
\begin{tabular}{|l|c|l|}
Facility &	Depth (mwe)&		Suite of detectors and technology \\ \hline\hline
Japan  &	Surface&	1 HPGe (with active veto) managed by KamLAND \\ 
& &\ \ \ \  currently 100\%, but may consider use by others \\ 
& & 1 HePG managed by CANDLES in Osaka (sea level) \\ 
& & \ \ \ \  currently 90\%, with 10\% use by others \\ \hline
Kamioka    &
	2700  &	Each experimental group has their own devices for  \\ 
\ \ Observatory  & &\ \  screening and assay, but will consider use by others \\ 
Japan & & 1 HPGe managed by KamLAND, currently 100\% \\
& &1 HPGe under construction by KamLAND: 100\% \\
& & 1 HPGe under construction by CANDLES: 100\%\\
& & 3 HPGe (2 p-type, 1 n-type)100\% SuperK and XMASS  \\ 
 & & Underground ICP-MS   and  API-MS   \\ 
 & &   \ \ \ \          managed by SuperK and XMASS (100\%) \\ 
  & & Many radon detectors to measure radon emanation of \\ & & \ \ \ \ materials, managed by SuperK and XMASS  (100\%)  \\  \hline 
CanFranc (LSC) & 
	2450 & 	5 HPGe p-type 100\% usage by LSC \\ 
Spain & &  \ \ \ \      Outside collaboration possible \\
& & 2 HPGe p-type to be installed by end of summer 2013 \\ \hline
Boulby Mine  &   2805  & 
	1 HPGe 2kg p-type for UK groups in SuperNEMO, \\ 
England & & \ \ \ \  LZ, DRIFT (75\%), 25\% use by others. \\ 
& & 2 n-type LO-AX detectors for radiometric dating: \\
& & \ \ \ \ 80\% Boulby/Scottish Universities Environmental\\
& & \ \ \ \ Research Centre, 20\% use by others
  \\ & & 1 new broad energy p-type HPGe (mid-2014) \\
& &\ \ \ \  100\% dedicated to SuperNEMO, LZ, DRIFT \\ & &  Expanded assay capability with new lab construction \\ \hline	
STELLA  at LNGS  & 
	3800  &
	10 HPGe operated by INFN as a user facility \\ 
Gran Sasso & & 1 HPGe with 100\% usage by XENON and GERDA,\\ 
Italy  & & \ \ \ \ (DARWIN in future), radon mitigation underway\\ \hline
LSM (Modane) &
	4800 &	15 HPGe with 6  dedicated to material selection: \\ 
France & & \ \ \ \ 	2 detectors, 100\% usage by SuperNEMO \\ 
& & \ \ \ \	1 detector 100\% usage by EDELWEISS \\ 
& & \ \ \ \ 	3 detectors 100\% dedicated to experiments \\
& & \ \ \ \ \ \ \ \  installed in Modane \\ 
& & 2 detectors may be available to others at level of 5-10\%\\ \hline
\end{tabular}

\bigskip 

\caption{International assay resources~\cite{refone,reftwo}.}
\label{tab:two}
\end{center}
\end{table} 

\begin{table}[t]
\begin{center}
\begin{tabular}{|l|c|l|}
Facility &	Depth (mwe)&		Suite of detectors and technology \\ \hline\hline
SNOLAB  &
	6010  &	1 PGT coax HPGe 54\% usage by Canadian based\\ 
 Canada & & \ \ \ \  experiments, 34\% usage by U.S. based experiments, \\
& & \ \ \ \  12\% usage by SNOLAB \\ 
& & 1 Canberra well HPGe , 100\% by SNO+ and DEAP \\
& & 11 Electrostatic counters (alpha counters), 100\% usage\\ & & \ \ \ \  by EXO, in future SNO+, PICASSO and MiniCLEAN\\ 
& & 8 alpha-beta counters, 100\% usage by SNO+ \\ 
 & & \ \ \ \     available for other experiments on request \\ 
& & 1 Canberra coax HPGe (currently being refurbished)\\
& & The SNOLAB facilities are used by SNOLAB based \\ & & experiments, but can be negotiated during down time \\ \hline
CJPL (JinPing) &
	6800  &	1 HPGe managed by PandaX, 100\% for PandaX \\ 
China & & 1 HPGe managed by CDEX, 90\% usage by CDEX\\ 
& & 2 HPGe to be installed by end of 2013: ~ 70\% CDEX,  \\ & & \ \ \ \ $\sim$30\% availability reserved for others.\\ \hline 
\end{tabular}

\bigskip 

\caption{International assay resources (cont.) ~\cite{refone,reftwo}.}
\label{tab:three}
\end{center}
\end{table} 

The original vision for DUSEL included the same attention to assay infrastructure as is manifested in the investments of Europe, Canada, and Japan (and now China).  However, without such a National Underground Laboratory to concentrate screening in one location and to justify the expense, we may require a new model.  Over the last decade, the gap between the needs of rare event physics searches and available screening has widened considerably, since it has been left to individual experiments to propose individually what should be common to all. This has led to a shortage of screening infrastructure overall and uneven distribution of existing resources.

\subsection{Assay techniques required}

High-purity germanium gamma ray spectroscopy (HPGe) is a mature technology which serves as the prime tool for material selection. Moderate sensitivity to the ubiquitous background from radioisotopes of potassium and the uranium/thorium chains is routinely achieved using commercially available detectors. A large number of such moderate sensitivity, “production” screeners are needed to deal with the high throughput of samples expected over the next decade. To meet this need, additional HPGe stations should be located at the three U.S. underground sites, which already have the staffing and expertise to produce quick turn-around for multiple users. 

For the most sensitive applications, R\&D in ultra-sensitive HPGe technology should be pursued at several sites where expertise already exists.  With appropriate choice of low-background cryostat and shielding, multiple crystal coincidence detectors with pulse shape discrimination can reach $\sim$ 10$^{-12}$ g/g.  HPGe with thin windows and low thresholds are useful for extending background characterization into the x-ray region (useful for solar neutrino projects), understanding cosmogenic activation, and surface contamination.  

The signal from many trace isotopes can be enhanced using neutron activation analysis (NAA) to observe counts from short-lived isotopes produced by epithermal neutron capture.  Such enhanced signals do not need a shielded underground site to achieve ppt sensitivity and can be counted by the production screeners discussed above.  This also goes for pre-screening of unknown materials, which might contaminate more sensitive screeners.   

There is also a need for alpha and beta screening for contaminants that are not accompanied by gamma emission. $^{210}$Pb and its progeny do not have a penetrating signature and are deposited on all surfaces exposed to radon.  Such radon plate-out plagues all rare-events searches, since it creates a patina of contaminant that causes nuclear recoils, beta-emission, and alphas.  In addition, since Pb is often used in circuitry, its alpha-decay to $^{210}$Po can cause single-site upsets. Surface contaminants, such as $^{40}$K, and anthropogenic contaminants, like $^{125}$Sb and $^{137}$Cs, are also detectable by beta screening. R\&D into more sensitive counters should be supported. 

Radioisotope identification can be done using mass spectrometry on the surface.  While machines capable of probing to ppb levels are typically found in universities and commercial analytic labs, the ppt levels achieved by inductively-coupled plasma mass spectroscopy (ICPMS) are only realized at a few locations where radiochemical expertise and careful attention to quality assurance protocols are augmented by novel dissolution and digestion techniques which can process a wide variety of samples.  Since ICPMS provides an alternative to counting, it must be taken into account in any determination of future assay needs, as well as included in any centralized scheduling apparatus. 

Laser cooling techniques can be used to trap atoms, excite them to a metastable state, and then detect their fluorescence, thus determining abundances by directly counting atoms. AtomTrap Trace Analysis provides a fast turn-around method of measuring radioactive background from $^{85}$Kr and $^{39}$Ar to a few parts in 10$^{-14}$  and could be installed underground to screen user samples, as well as aid in the purification of Ar, Ne, Xe.

While an augmented suite of sensitive production screeners can provide the bulk of the assay, orders of magnitude improvement in sensitivity is required for some materials close to detectors and for active elements.  An ultra-sensitive whole body screener in a water tank at depth would provide an ultimate check on the total activity from all isotopes in the material, including short-lived isotopes which are impossible to detect chemically. The goal is bulk assay of large amounts of material at the $10^{-13}$--$10^{-14}$ g/g U/Th level.  Such designs have been explored in NUSL~\cite{refthree} and DUSEL~\cite{reffour}  reports and generally resemble the Borexino Counting Test Facility~\cite{reffive}  with a top-loading sample changer.

\subsection{R\&D space}

Reconfigurable space must be reserved underground for small scale experiments, storage, and background studies. Prototypes using new technologies for next-generation dark matter and neutrinoless double-beta decay experiments must be operated underground to establish background rejection and evaluate sensitivity.  While muons can be efficiently tagged on the surface, muon-induced neutrons where the parent muon does not intersect the active shield produce an irreducible background that makes it impossible to determine the level of rejection achieved by the new technique.   By the same rationale, new assay techniques also require underground R\&D space to evaluate their sensitivity before deployment.   

Underground space is not low-background unless it is also shielded against radiogenic backgrounds from the rock or the shotcrete which covers the rock.  Most users need a combination of high-Z material for gamma reduction, hydrogenous material for neutron absorption, and copper liners for the cleanest inner shell.  Since shielding is often the most expensive element in the experiment, the shielding should remain underground to be re-used as part of the site infrastructure. This also has benefits with respect to long-term reduction in cosmic activation.  If the hydrogenous material is thick enough and pure enough, it is sufficient on its own.  This implies that the most cost-effective shielding for multiple users would be a large water tank surrounding the R\&D space, sharing water purification infrastructure already needed for the neighboring ton-scale dark matter experiments. 

A better quantification of backgrounds underground is important to all experiments and should be part of the supplied infrastructure.  For example, each site should be able to provide spectral and flux characterization of their muon, gamma, and neutron backgrounds to the experiments located there.  Compilation of these data over multiple geologies and depths provides the means to better model the physics, as well as take the burden from individual experiments.  Although the rock overburden reduces the flux of cosmic-ray muons by several orders of magnitude, a residual flux of highly penetrating muons creates difficult-to-shield high-energy neutrons.  The neutrons themselves can imitate WIMP interactions, and gammas from their inelastic scatters produce gamma background for neutrinoless double-beta decay experiments.  Data on the flux, spectrum, and composition of such cosmogenic neutrons underground is therefore vital, but since the rate is so low, experiments must rely on extensive Monte Carlo simulations for background estimates/subtraction, shielding design, and veto efficiencies.  Large vats of gadolinium or boron-loaded water or liquid scintillator can provide the data to compare to simulation and improve the physics models.  These can serve multiple purposes, including shielding for experiments and R\&D, as well as neutrino and homeland security applications.

\subsection{Strategic Infrastructure Planning}

Immediate investment in assay infrastructure is required to ensure the successful deployment of the next-generation dark matter, neutrino, and neutrinoless double-beta decay experiments.  The availability of screeners worldwide is not sufficient for the suite of experiments currently approved and proposed. This is an international shortage, so sending samples overseas for assay is not the solution, although experiments with significant international collaborators will certainly employ sensitive screeners at European and Asian labs for a subset of samples. Since sensitive counting techniques require overburdens matched to existing U.S. underground facilities, space in U.S. laboratories should be reserved for this function.  There is enough space for future expansion of assay capabilities, and for new small-scale experiments designed to test new technologies or expand sensitivity to new WIMP regimes, as long as the full suite of existing U.S. laboratories is exploited.   This would also take advantage of considerable past investment, existing infrastructure, and expert personnel. Using U.S. sites will also minimize surface activation during shipping and contamination during handling. 

Increased U.S. assay capability is best achieved by organizing the existing underground facilities under a single umbrella organization with representatives from those sites, which could then be funded as a consortium or network, rather than as a competition between experiments.  Such an entity can then propose increased investment in capability as a validated representative of the larger community, including centralized screening at the deep U.S. site to take advantage of co-location and unique smaller efforts at other locations.  In the meantime, inter-site scheduling tools can be developed to achieve improved efficiency with existing screeners. The same organization can manage R\&D space underground, including scheduling use, underground storage of cosmogenically-activated materials, and multi-user shielding. The larger community thus served would form a base for innovative funding schemes which include industry partners.  

A staged transition from an experiment-specific model to a multi-site user coordinated  network is in line with current budgetary constraints, yet would retain capabilities that have been built up over many years in the service of individual projects. Such organization will promote rapid and broad dissemination of research results.  When combined with access to assay and a database of previously assayed materials, this will significantly increase scientific reach and reduce risk for all rare event searches underground.

\section{Conclusions}

In summary, a substantial expansion in non-U.S. underground capabilities is very likely to occur by the end of this decade. The U.S. has leading roles in many of the future dark matter, neutrinoless double-beta-decay, and neutrino experiments. It is critical that U.S. scientists continue to be supported well enough to take full advantage of international and domestic underground facilities housing these experiments. Improved coordination of the underground facilities (including domestic infrastructure) is required to maintain this leading role. An open access policy for all major underground facilities is needed. As the scale (and thus cost) of underground experiments grows, it will become more important to foster open, competitive access as much as possible. The best way for the governments to support the international system of underground experiments is for each major country (or region) to support at least one major underground laboratory capable of hosting the forefront experiments. It is not clear whether it would be possible to sustain this international support if one country chose to take a major role in the research without supporting any facility. 

Our conclusions related to the upcoming U.S. planning process are:
\begin{enumerate}
\item 	Locate LBNE underground to realize its full science potential. This step would also provide a natural base for additional domestic underground capabilities at SURF in the future.
\item	The U.S. has leading roles in many of the future dark matter, neutrinoless double beta decay and neutrino experiments. 
\item	More coordination and planning of underground facilities (overseas and domestic) is required to maintain this leading role, including use of existing U.S. infrastructure and closer coordination with SNOLAB as the deepest North American Lab. 
\item 	Maintaining an underground facility that can be expanded to house the largest dark matter and neutrinoless double beta decay experiments would guarantee the ability of the U.S. to continue its strong role in the worldwide program of underground physics.
\end{enumerate}




\end{document}